\documentclass{emulateapj}

\shorttitle{The $M_{\rm bh}$-$\sigma$ diagram}
\shortauthors{Alister W.\ Graham}
\slugcomment{accepted for publication in ApJ, March 16, 2009}

\begin{document}

\title{The $M_{\rm bh}$-$\sigma$ diagram, and the
  offset nature of barred active galaxies}

\author{Alister W.\ Graham\altaffilmark{1} and I-hui Li}
\affil{Centre for Astrophysics and Supercomputing, Swinburne University
of Technology, Hawthorn, Victoria 3122, Australia.}
\altaffiltext{1}{Corresponding Author: AGraham@swin.edu.au}

\begin{abstract} 

From a sample of 50 predominantly inactive galaxies with direct supermassive
black hole mass measurements, it has recently been established that barred
galaxies tend to reside rightward of the $M_{\rm bh}$-$\sigma$ relation
defined by non-barred galaxies.
Either black holes in barred galaxies tend to be anaemic or the central
velocity dispersions in these galaxies have a tendency to be elevated by the
presence of the bar.
The latter option is in accord with studies connecting larger velocity
dispersions in galaxies with old bars, while the former scenario is at odds
with the observation that barred galaxies do not deviate from the $M_{\rm
 bh}$-luminosity relation.
Using a sample of 88 galaxies with active galactic nuclei, whose supermassive
black hole masses have been estimated from their associated emission lines,
we reveal for the first time that they also display this same general behavior
in the $M_{\rm bh}$-$\sigma$ diagram depending on the presence of a bar or not.
A new symmetrical and non-symmetrical ``barless'' $M_{\rm bh}$-$\sigma$
relation is derived using 82 non-barred galaxies.  The barred galaxies are shown to
reside on or up to $\sim$1 dex below this relation.
This may explain why narrow-line Seyfert 1 galaxies appear offset from the 
``barless'' $M_{\rm bh}$-$\sigma$ relation, and has far reaching implications
given that over half of the disk galaxy population are barred. 

\end{abstract}

\keywords{ 
black hole physics --- 
galaxies: active --- 
galaxies: nuclei --- 
galaxies: Seyfert --- 
galaxies: structure }

\section{Introduction}

Black holes fascinate astronomers and the general public alike due to their
extreme physical conditions.  How the supermassive variety came to be at the
centres of most, if not all, massive galaxies has been the focus of increased 
research over recent years.  Valuable clues are thought to reside
within the observed scaling laws involving the mass of the central black hole,
$M_{\rm bh}$, and various properties of the host galaxy (e.g., Ferrarese \&
Ford 2005 and references therein).  Indeed, the low level of scatter within
these relations suggests a direct physical connection between the black hole
and the host galaxy; although which parameters are doing the driving, or are
merely along for the ride, remains unclear (Novak et al.\ 2006).

One well known scaling law involves the velocity dispersion, $\sigma$, of the
host galaxy (Ferrarese \& Merritt 2000; Gebhardt et al.\ 2000).  With
parameter measurement errors initially of comparable size to the total
root mean square (r.m.s.) 
scatter about the $M_{\rm bh}$-$\sigma$ relation, this scaling law was
heralded as a potential fundamental relation with zero intrinsic scatter
(Ferrarese \& Merritt 2000).  If correct, it would imply that the host
galaxy's velocity dispersion controls the growth of supermassive black holes (SMBHs),
and is the physical property which theorists should concentrate their efforts
on.  The situation did however become more interesting, if not complicated,
when Graham (2007; 2008a,b) and Hu (2008) revealed the presence of
substructure within the $M_{\rm bh}$-$\sigma$ plane.  They discovered that the
location of a galaxy/black hole pair in this diagram depends on whether or
not the galaxy is barred (or has a ``pseudo-bulge'' in Hu's presentation).
Moreover, the offset of barred galaxies from the ``barless $M_{\rm
  bh}$-$\sigma$ relation'' can be up to $\sim$1 dex in the $\log M_{\rm bh}$
direction.

Graham (2008a) suggested that the radial orbits of stars in bars may lead to
an enhanced velocity dispersion measurement when one is looking down (some
component of) the length of a bar.  In addition, bars which have evolved for a
sufficient amount of time are expected to increase their galaxy's central
velocity dispersion, even in face on galaxies (Gadotti \& de Souza 2005, and
references therein).  Indeed, bars with an older stellar population have been
observed to possess higher velocity dispersions than younger bars (Perez et
al.\ 2009).  This will contribute to, if not explain, the discrepant behavior
of barred galaxies in the $M_{\rm bh}$-$\sigma$ diagram.  Alternatively or
additionally, the supermassive black holes in barred galaxies may be
relatively malnourished compared to those in barless galaxies.  That is,
perhaps they are still being fuelled/fed by the bar\footnote{We note that it 
is not clear whether bars funnel gas all the way to the center of their 
galaxies, e.g. Forbes et al.\ 1994.} 
and as such are yet to
reach the barless $M_{\rm bh}$-$\sigma$ relation (Mathur \& Grupe 2005a; Zhou
et al.\ 2006). However, if this were true,
then the barred galaxies would also be outliers in the $M_{\rm bh}$-luminosity
diagram, yet data to date suggests that they are not (Graham 2008a; Bentz et
al.\ 2009).

Ferrarese et al.\ (2001, see also Wang \& Lu 2001) 
have shown that black holes in AGN roughly abide by
the $M_{\rm bh}$-$\sigma$ relation defined by quiescent galaxies.  In this
paper we explore the demographics within the $M_{\rm bh}$-$\sigma$ plane by
including a sample of 88 active galactic nuclei (AGN).  Specifically, we
address the question of whether or not, at a fixed black hole mass, do barred
galaxies (with AGN) have a tendency to posses greater velocity dispersions
than non-barred galaxies (with AGN).  In Section~2 we present the galaxy
sample, and our method of image analysis to determine if a bar may be present.
Images for the full AGN galaxy sample are available electronically.  In Section~3
we show the location of the barred and non-barred galaxies within the $M_{\rm
  bh}$-$\sigma$ diagram.  Galaxies with AGN are revealed, for the first time, to follow the
same distributions as inactive barred and unbarred galaxies.  That is, barred
AGN reside up to 1 dex in the $\log M_{\rm bh}$ direction below the barless
$M_{\rm bh}$-$\sigma$ relation.  Our main conclusions are presented in Section
4.

\section{Data}

Broad emission-line reverberation-mapping data was used by Peterson et al.\
(2004) to acquire SMBH virial masses for 35 AGN, which were subsequently used
to calibrate a number of scaling relations for estimating SMBH masses in other
AGN (e.g., Greene \& Ho 2005; Kaspi et al.\ 2005).  Here we have used Greene
\& Ho's (2006, their Table~1) useful list of 88 local active galaxies with
stellar velocity dispersion measurements and estimated black hole
masses.\footnote{Although the normalization (from the scaling factor $f$
  pertaining to the broad line region's geometry and kinematics) of these SMBH
  masses is tied to the $M_{\rm bh}$-$\sigma$ relation of inactive galaxies
  (Onken et al.\ 2004), ``circular rationale'' should not be the cause 
  of the barred and unbarred AGN potentially occupying the same distribution
  and locus as barred and unbarred inactive galaxies in the $M_{\rm
    bh}$-$\sigma$ plane.}  Such estimates are typically considered accurate to
within a factor of 3-4 (e.g., Krolik 2001; Metzroth et al.\ 2006; Vestergaard
\& Peterson 2006).  We then determined if a bar is present or not in these
galaxies.
% 
% A Hubble constant of XXX km s$^{-1}$ Mpc$^{-1}$ was used to convert their
% recessional velocities into distances and thus determine the black hole masses.

\begin{figure}
\includegraphics[angle=90,scale=0.81]{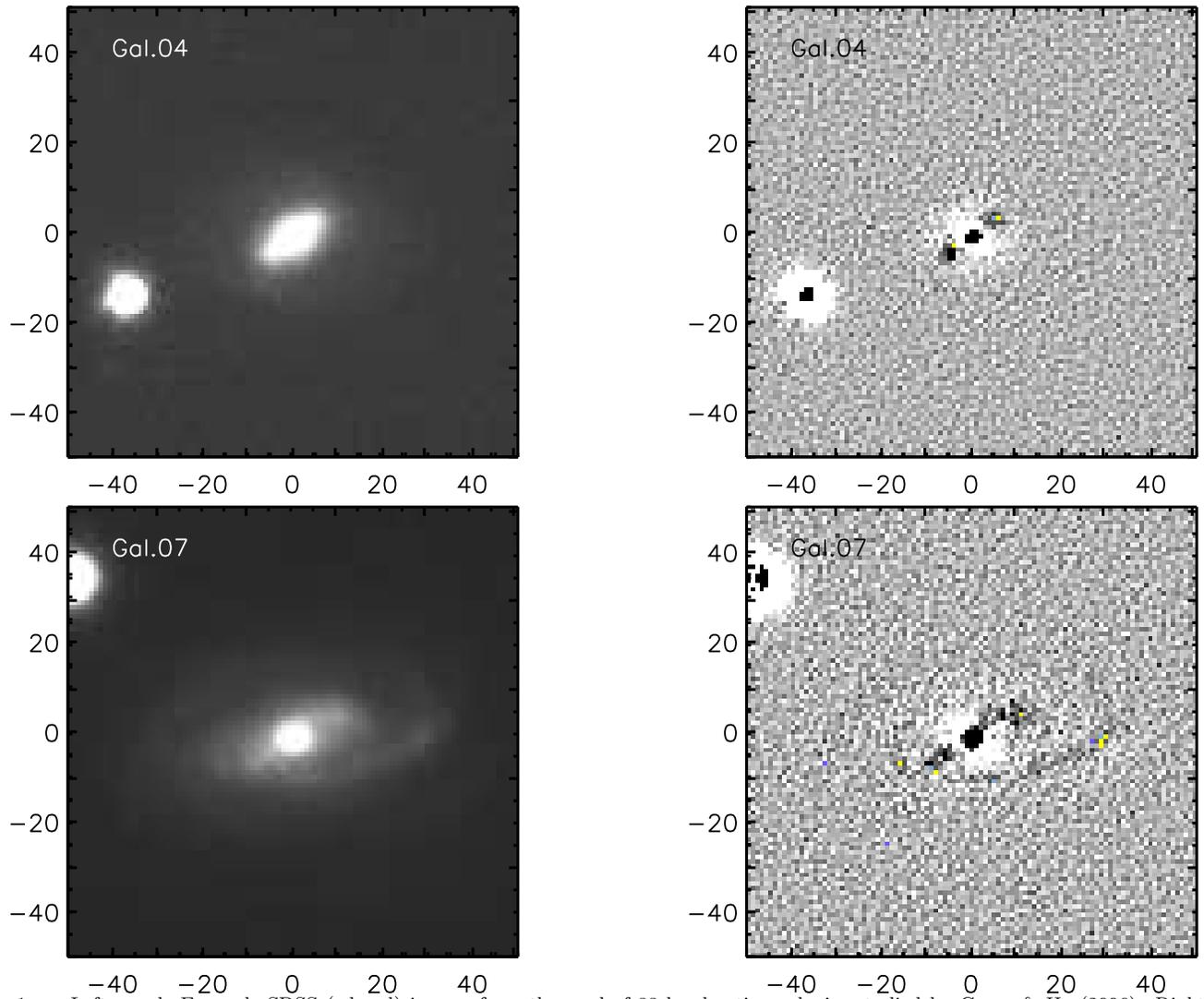}
\caption{
Left panel: Example SDSS (r-band) images from the pool of 88 local active
galaxies studied by Green \& Ho (2006).  Right panel: Unsharp masking
helps to reveal the presence of bars.  
The axis tick marks denote pixels, which correspond to a size of 0.396 arcseconds.  
Image pairs for all 88 galaxies are available via the electronic version of this paper. 
Note: The positive/negative (white/black) has been reversed in the residual images.
}
\label{Fig1}
\end{figure}

\begin{figure}
\includegraphics[angle=270,scale=0.9]{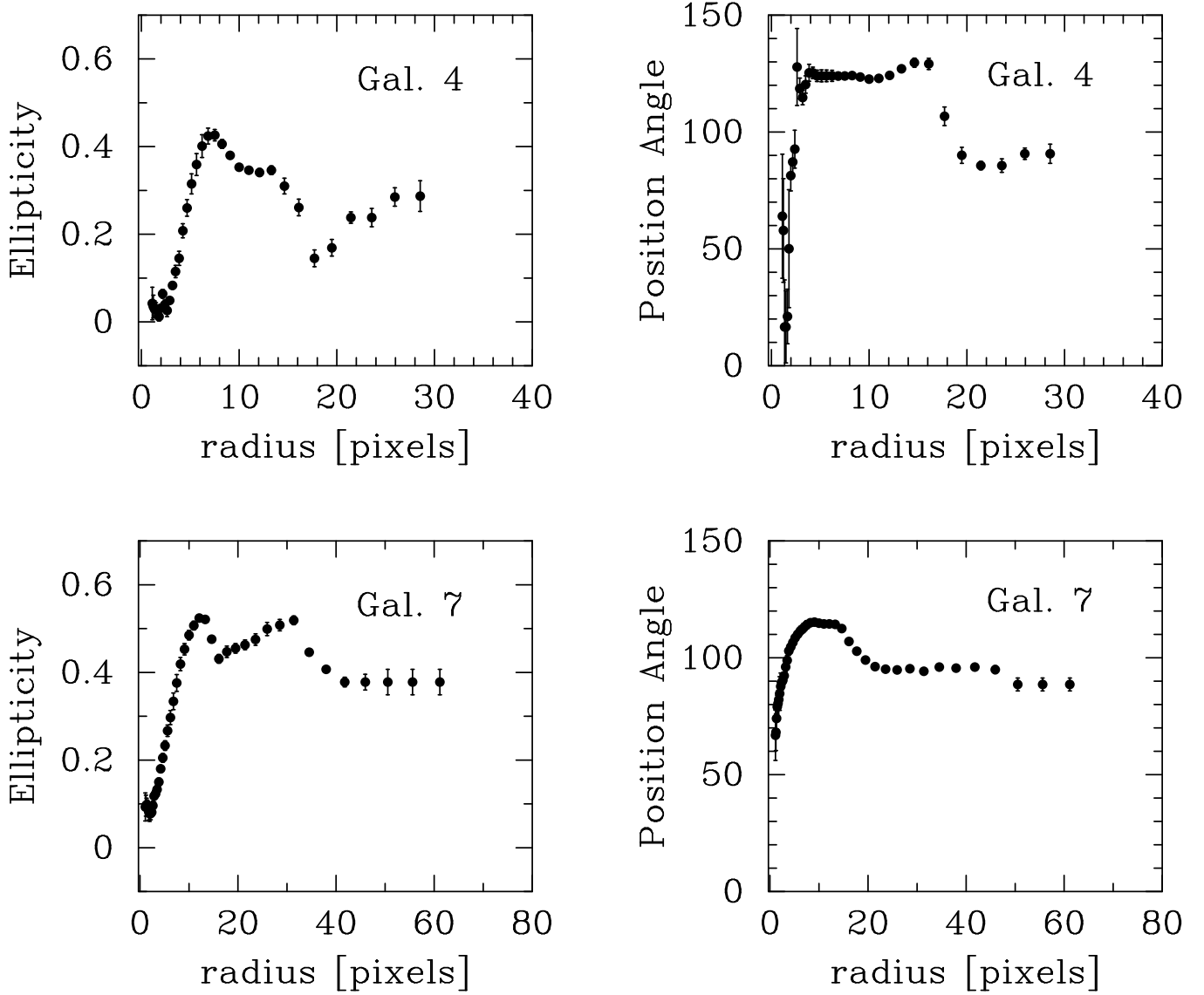}
\caption{
Ellipticity and position angle profiles for the galaxies shown in Figure~\ref{Fig1}
}
\label{Fig2}
\end{figure}

Images for the 88 AGN have predominantly been taken from the Sloan Digital Sky Survey
(SDSS; Abazajian et al.\ 2008 and references therein), with the remaining images
obtained via the NASA Extragalactic Database (NED).  While a visual
inspection of the images can, and often does, reveal bars in many
galaxies, it also leaves some ambiguous cases.  To help address these we have
subtracted from the original image a number of smoothed representations.
Sometimes this smoothed version was an elliptical object with the same median
ellipticity and position angle as the galaxy, while other times it was a
Gaussian smoothed version of the original image.  This common technique (e.g.,
Jerjen et al.\ 2000; Yuan et al.\ 2001, Erwin \& Sparke 2002) created a set of
residual images which can more clearly reveal non-symmetrical structures such
as stellar bars.
A couple of example images are shown in Figure~\ref{Fig1}, and a complete set of
images for all 88 galaxies is available electronically.  In addition we used
the IRAF task ELLIPSE to generate ellipticity and position angle profiles,
which were found to be helpful in identifying/confirming the presence of bars.
This was typically signalled by a high ellipticity over the inner regions once
the bar light starts to dominate; the ellipticity profile would then decline with increasing radius
while often accompanied by a changing position angle as the outer disk comes
into dominance (Figure~\ref{Fig2}).

In Table~\ref{Tab1} we have identified which galaxies appear to have bars or
not.  For some galaxies we were left uncertain as to whether what appeared to be a bar
was real, and we have assigned these with the status ``maybe''.  
A ``no'' signals that we saw no evidence for a bar.  While this system may sound
obvious, it is worth clarifying that poorly resolved or edge-on 
systems in which a bar might exist but for which we saw no evidence are
designated as ``no'' rather than ``maybe''. 
Given our suspicion that the outlying galaxies in the $M_{\rm bh}$-$\sigma$
diagram are barred, a stronger test 
of our hypothesis would therefore involve such an economical assignment of bars. 
We have however denoted five galaxies (NGC~3227, 3516, 3783, 4151, and 7469) to be barred
given their designation as such by others, according to NED, 
although this was not clear from the SDSS images. 
From the 88 AGN we were able to positively (tentatively) identify 26 (16)
barred galaxies, giving a {\it total} fraction of 42/88 which is roughly half
of the sample.  We do however note that for twenty small and faint galaxies
with $\sigma < 85$ km s$^{-1}$, it was hard to detect the presence of a
bar and all but one have been tabulated as having no (detectable) bar.

\clearpage

\begin{deluxetable}{rlrrlll}
\tablewidth{400pt}
\tablecaption{Galaxy Properties \label{Tab1}}
\tablehead{
\colhead{Gal.\ Id.} & \colhead{CDS Name} & \colhead{R.A.[Deg]} & \colhead{Dec.\ [Deg]} & \colhead{Classification} & \colhead{$\epsilon$} & \colhead{Bar} \\ % spiral
(1) &   (2)                        &     (3)      &  (4)        &  (5) &  (6)     & (7)     \\ % (8)
}
\startdata
 1  &   SDSS~J000805.62+145023.4   &     2.02333  &  14.83972   &  Sy1           &  0.41   & yes    \\ % yes  
 2  &   SDSS~J004236.86-104921.8   &    10.65417  & -10.81000   &  Sy1.5         &  0.18   & maybe  \\ % no    
 3  &   SDSS~J010712.03+140844.9   &    16.80000  &  14.14583   &  NLSy1?        &  0.05   & no     \\ % no    
 4  &   SDSS~J011703.58+000027.3   &    19.26500  &   0.00750   &  NLSy1         &  0.29   & yes    \\ % no?   
 5  &   SDSS~J020459.25-080816.0   &    31.24687  &  -8.13778   &  ...           &  0.45   & yes    \\ % no?   
 6  &   SDSS~J020615.99-001729.1   &    31.56708  &  -0.29139   &  S0;merger?    &  0.50   & no     \\ % no?   
 7  &   SDSS~J021011.49-090335.5   &    32.54787  &  -9.05986   &  Sa            &  0.38   & yes    \\ % yes   
 8  &   SDSS~J021257.59+140610.1   &    33.23996  &  14.10281   &  BLAGN         &  0.40   & no     \\ % yes   
 9  &                    Mrk~590   &    33.64000  &  -0.76667   &  SA(s)a        &  0.12   & no     \\ % maybe 
10  &   SDSS~J024912.86-081525.6   &    42.30375  &  -8.25722   &  BLAGN         &  0.25   & no     \\ % no    
11  &   SDSS~J032515.59+003408.4   &    51.31500  &   0.56889   &  ...           &  0.10   & no     \\ % no    
12  &   SDSS~J033013.26-053236.0   &    52.55500  &  -5.54333   &  Sb            &  0.40   & maybe  \\ % yes   
13  &                     3C~120   &    68.29625  &   5.35444   &  S0;LPG;BLRG;  &  0.40   & maybe  \\ % maybe?
14  &                    Akn~120   &    79.04792  &  -0.15028   &  Sb/pec        &  0.99   & no     \\ % maybe?
15  &                     Mrk~79   &   115.63708  &  49.80917   &  SBb           &  0.50   & yes    \\ % yes   
16  &   SDSS~J075057.25+353037.5   &   117.73854  &  35.51042   &  ...           &  0.99   & no     \\ % no    
17  &   SDSS~J080243.39+310403.3   &   120.68083  &  31.06750   &  AGN           &  0.18   & no     \\ % no    
18  &   SDSS~J080538.66+261005.4   &   121.41125  &  26.16806   &  SB(s)bc       &  0.30   & yes    \\ % yes   
19  &   SDSS~J080907.58+441641.4   &   122.28167  &  44.27806   &  Sy            &  0.20   & no     \\ % no    
20  &   SDSS~J082510.23+375919.7   &   126.29250  &  37.98889   &  ...           &  0.65   & maybe  \\ % yes   
21  &   SDSS~J082912.67+500652.3   &   127.30292  &  50.11444   &  BLAGN         &  0.10   & no     \\ % no    
22  &   SDSS~J083202.16+461425.7   &   128.00500  &  46.24389   &  Sy1           &  0.20   & yes  \\   % yes   
23  &   SDSS~J083949.64+484701.4   &   129.95750  &  48.78361   &  Sy1           &  0.25   & maybe  \\ % yes   
24  &   SDSS~J085554.27+005110.9   &   133.97612  &   0.85303   &  Sy1           &  0.40   & no     \\ % yes   
25  &   SDSS~J092438.88+560746.9   &   141.16375  &  56.12861   &  NLSy1         &  0.50   & yes    \\ % yes   
26  &                    Mrk~110   &   141.30375  &  52.28639   &  Pair?         &  0.99   & no     \\ % yes   
27  &   SDSS~J093259.60+040506.0   &   143.24833  &   4.08500   &  ...           &  0.03   & maybe  \\ % no?   
28  &   SDSS~J093812.26+074340.0   &   144.55125  &   7.72778   &  ...           &  0.20   & no     \\ % no    
29  &   SDSS~J094838.42+403043.7   &   147.16000  &  40.51222   &  Sy            &  0.50   & yes    \\ % yes   
30  &   SDSS~J101108.40+002908.7   &   152.78500  &   0.48583   &  ...           &  0.15   & no     \\ % no    
31  &   SDSS~J101627.32-000714.5   &   154.11375  &  -0.12056   &  ...           &  0.35   & no     \\ % maybe?
32  &   SDSS~J101912.57+635802.7   &   154.80250  &  63.96750   &  E;Sy1.5       &  0.40   & yes  \\   % no?   
33  &   SDSS~J102044.43+013048.4   &   155.18512  &   1.51344   &  BLAGN         &  0.10   & no  \\    % maybe 
34  &                   NGC~3227   &   155.87750  &  19.86500   &  SAB(s)        &  0.60   & yes    \\ % yes   
35  &   SDSS~J110640.20+051905.6   &   166.66750  &   5.31822   &  ...           &  0.25   & yes    \\ % yes   
36  &                   NGC~3516   &   166.69833  &  72.56889   &  (R)SB(s)0     &  0.27   & yes    \\ %  no?  
37  &   SDSS~J112536.16+542257.1   &   171.40083  &  54.38167   &  S0            &  0.55   & no     \\ % maybe?
38  &   SDSS~J112841.00+575006.5   &   172.17083  &  57.83514   &  Sy2           &  0.34   & yes    \\ % yes   
39  &                   NGC~3783   &   174.75750  & -37.73861   &  (R')SB(r)a    &  0.15   & yes    \\ % yes   
40  &                     POX~52   &   180.73708  & -20.93417   &  ...           &  0.45   & no     \\ % no    
41  &   SDSS~J120257.81+045045.0   &   180.74083  &   4.84583   &  (R')SA(s)c:   &  0.50   & maybe  \\ % yes   
42  &                   NGC~4051   &   180.79000  &  44.53139   &  SAB(rs)bc     &  0.55   & yes    \\ % yes   
43  &   SDSS~J120556.01+495956.1   &   181.48337  &  49.99892   &  ...           &  0.18   & no     \\ % no    
44  &                   NGC~4151   &   182.63625  &  39.40556   &  (R')SAB(rs)ab &  0.50   & yes    \\ % yes   
45  &   SDSS~J121607.09+504930.0   &   184.02958  &  50.82500   &  SBb?          &  0.66   & yes    \\ % yes   
46  &   SDSS~J121754.97+583935.6   &   184.47917  &  58.66000   &  compact       &  0.21   & no     \\ % no    
47  &   SDSS~J122324.13+024044.4   &   185.85042  &   2.67917   &  E?            &  0.35   & no     \\ % no    
48  &                   NGC~4395   &   186.45375  &  33.54667   &  SA(s)m;LINER  &  0.15   & no     \\ % no    
49  &   SDSS~J123237.48+662452.3   &   188.15583  &  66.41444   &  BLAGN         &  0.40   & no     \\ % maybe 
50  &   SDSS~J124035.81-002919.4   &   190.14917  &  -0.48861   &  AGN           &  0.05   & no     \\ % no    
51  &   SDSS~J125055.28-015556.6   &   192.73042  &  -1.93250   &  AGN           &  0.10   & no     \\ % no    
52  &   SDSS~J130620.97+531823.1   &   196.58750  &  53.30639   &  ...           &  0.25   & no    \\  % no    
53  &   SDSS~J131305.80+012755.9   &   198.27417  &   1.46556   &  BLAGN         &  0.08   & maybe  \\ % maybe 
54  &   SDSS~J132249.21+545528.2   &   200.70504  &  54.92450   &  Sy1.5         &  0.40   & no     \\ % maybe 
55  &   SDSS~J132340.31-012749.2   &   200.91796  &  -1.46367   &  Sy1.5         &  0.10   & no     \\ % maybe 
56  &   SDSS~J134952.84+020445.1   &   207.47000  &   2.07917   &  Sy1           &  0.50   & maybe  \\ % yes   
57  &                    Mrk~279   &   208.26458  &  69.30806   &  S0            &  0.34   & no     \\ % maybe?
58  &   SDSS~J140018.42+050242.2   &   210.07667  &   5.04500   &  ...           &  0.28   & maybe  \\ % yes   
59  &   SDSS~J140514.87-025901.2   &   211.31196  &  -2.98367   &  ...           &  0.55   & yes    \\ % yes   
60  &   SDSS~J141630.81+013708.0   &   214.12837  &   1.61889   &  BLAGN         &  0.30   & maybe  \\ % yes   
61  &                   NGC~5548   &   214.49875  &  25.13694   &  (R')SA(s)0/a  &  0.20   & no     \\ % yes   
62  &   SDSS~J143450.62+033842.5   &   218.71125  &   3.64444   &  Sc            &  0.12   & maybe  \\ % yes   
63  &   SDSS~J143452.45+483942.7   &   218.71875  &  48.66194   &  SB(s)0/a?     &  0.10   & no   \\   % no    
64  &                    Mrk~817   &   219.09250  &  58.79417   &  SBc           &  0.10   & yes    \\ % yes   
65  &   SDSS~J144629.97+500130.5   &   221.62500  &  50.02528   &  ...           &  0.22   & no     \\ % yes   
66  &   SDSS~J145706.80+494008.4   &   224.27833  &  49.66917   &  SB(s)b        &  0.10   & maybe  \\ % yes   
67  &   SDSS~J145901.35+611353.5   &   224.75542  &  61.23139   &  S             &  0.38   & yes    \\ % yes   
68  &   SDSS~J150556.55+034226.3   &   226.48562  &   3.70731   &  Sa/b          &  0.50   & maybe  \\ % yes   
69  &   SDSS~J150745.00+512710.2   &   226.93750  &  51.45278   &  Sy1           &  0.65   & yes    \\ % maybe 
70  &   SDSS~J150853.95-001148.9   &   227.22479  &  -0.19692   &  Compact       &  0.20   & no     \\ % no    
\enddata
\end{deluxetable}

\clearpage
\setcounter{table}{0}

\begin{deluxetable}{rlrrlll}
\tablewidth{400pt}
\tablecaption{Galaxy Properties {\it cont.}}
\tablehead{
\colhead{ } & \colhead{CDS Name} & \colhead{R.A.[Deg]} & \colhead{Dec.\ [Deg]} & \colhead{Classification} & \colhead{$\epsilon$} & \colhead{Bar} \\ % spiral
(1) &   (2)                        &     (3)      &  (4)        &  (5)           &  (6)     & (7)   \\  % (8)
}
\startdata
71  &   SDSS~J152515.18+601409.0   &   231.31325  &  60.25833   &  Sy2           &  0.08   & maybe  \\  % no    
72  &   SDSS~J155417.43+323837.8   &   238.57262  &  32.64383   &  AGN           &  0.15   & maybe  \\  % no    
73  &   SDSS~J161156.31+521116.8   &   242.98458  &  52.18806   &  BLAGN         &  0.30   & no     \\  % no?   
74  &   SDSS~J161951.31+405847.2   &   244.96375  &  40.97944   &  SBab          &  0.60   & yes    \\  % yes   
75  &   SDSS~J170246.09+602818.9   &   255.69208  &  60.47194   &  ...           &  0.16   & no     \\  % no    
76  &   SDSS~J170328.96+614109.9   &   255.87083  &  61.68611   &  BLAGN         &  0.99   & no     \\  % no    
77  &   SDSS~J171550.49+593548.7   &   258.96037  &  59.59686   &  BLAGN         &  0.50   & yes    \\  % yes   
78  &   SDSS~J172759.15+542147.0   &   261.99625  &  54.36306   &  NLSy1?        &  0.15   & no     \\  % no    
79  &                   3C~390.3   &   280.53750  &  79.77139   &  BLRG          &  0.20   & no     \\  % no    
80  &   SDSS~J212401.90-002158.7   &   321.00792  &  -0.36603   &  AGN           &  0.99   & no     \\  % no    
81  &   SDSS~J215658.30+110343.1   &   329.24292  &  11.06194   &  Sy            &  0.30   & yes    \\  % yes   
82  &   SDSS~J222435.29-001103.8   &   336.14708  &  -0.18444   &  ...           &  0.15   & no     \\  % no    
83  &   SDSS~J223000.37-094622.1   &   337.50154  &  -9.77281   &  ...           &  0.30   & no     \\  % no?   
84  &                   NGC~7469   &   345.81583  &   8.87389   &  (R')SAB(rs)a  &  0.33   & yes    \\  % yes   
85  &   SDSS~J232159.06+000738.8   &   350.49625  &   0.12750   &  ...           &  0.35   & no     \\  % maybe?
86  &   SDSS~J232721.96+152437.3   &   351.84167  &  15.41028   &  Sy1           &  0.25   & no     \\  % no?   
87  &   SDSS~J233837.10-002810.3   &   354.65458  &  -0.46944   &  NLAGN         &  0.99   & no     \\  % no    
88  &   SDSS~J235128.78+155259.0   &   357.86992  &  15.88306   &  AGN           &  0.45   & no     \\  % maybe?
\enddata
\tablecomments{
Col.\ (1): Running galaxy identification number. 
Col.\ (2): Galaxy name from the Centre de Donn\'ees astronomiques de
Strasbourg (CDS)\footnote{http://cdsweb.u-strasbg.fr}. 
Col.\ (3): Right Ascension.
Col.\ (4): Declination. 
Col.\ (5): Classification in the NASA Extragalactic Database (NED)\footnote{http://nedwww.ipac.caltech.edu}.
Col.\ (6): (IRAF/ELLIPSE)-determined ellipticity, $\epsilon$, of the outer
isophotes --- indicative of the galaxy inclination for the disk galaxies. 
Col.\ (7): Whether or not a bar was detected.
Black hole masses and velocity dispersions are tabulated in Greene \& Ho (2006,
their Table~1). 
%
% Table 1 is available in its entirety via the link to the machine-readable version above.
}
\end{deluxetable}
\clearpage

\section{Results}

Figure~\ref{Fig3} reveals the location of the 88 galaxies with AGN in the $M_{\rm
  bh}$-$\sigma$ plane.  For clarity we have not plotted error bars here, but
they can be seen in Graham (2008b).

\subsection{Inactive galaxies}\label{Inactive}

To set the context, shown in Figure~\ref{Fig3}a) is a local sample of 50 predominantly
inactive galaxies for which direct supermassive black hole mass measurements
have been catalogued (Graham 2008b).  Of these 50 galaxies, 28
are disk galaxies, and 14 are barred.  That is, half of the disk
galaxies are barred, which is in fair agreement with optical studies of the
bar fraction in spiral galaxies.  For example, Knapen et al.\ (2000, see also
Marinova \& Jogee 2007, and Hern\'andez-Toledo et al.\ 2008; and 
Weinzirl et al.\ 2009) have reported that
$\sim$60\% of disk galaxies have bars, although it may be as high
as 75\% (e.g., Eskridge et al.\ 2000). 
% xxx Whyte et al 2002, Block et al. 2002, Buta et al 2004).
% 
These barred galaxies 
are denoted by crosses in Figure~\ref{Fig3}a) and they clearly display the reported tendency
to reside on or below the $M_{\rm bh}$-$\sigma$ relation defined by the
non-barred galaxies.  The ``barless $M_{\rm bh}$-$\sigma$ relation'' shown
in Figure~\ref{Fig3}a by the solid straight line was obtained by Graham
(2008b) using the (symmetrical) bisector linear regression routine BCES from
Akritas \& Bershady (1996), and assuming a 10\% uncertainty on the velocity
dispersion values. The relation is such that 
$\log(M_{\rm bh}/M_{\odot}) =
(8.25\pm0.05) + (4.39\pm0.32)\log [\sigma/200\, {\rm km\, s}^{-1}]$.  
The dashed lines in this panel reflect the $1\sigma$ uncertainty on the slope and
intercept of this relation, while the shaded area expands this boundary by
0.33 dex (the r.m.s.\ scatter in the $\log M_{\rm bh}$ direction about
the $M_{\rm bh}$-$\sigma$ relation for the non-barred galaxies). 

Given concerns from different corners about potential biases with some
regression techniques, we have checked the above relation using two additional codes.
First, we have used Tremaine et al.'s (2002) modified version of the routine
FITEXY (Press et al.\ 1992, their Section~15.3) to construct two relations: 
one which minimised the scatter in the $\log M_{\rm bh}$ direction and another
which minimised the scatter in the $\log \sigma$ direction (see Novak et al.\ 2006).
Averaging these two relations gives a slope and intercept of (4.17 + 4.67)/2 = 4.42 
and (8.26 + 8.26)/2 = 8.26, respectively. 
The second code which we have used is an IDL routine from Kelly (2007) which
employs a Bayesian method to account for measurement errors.  The median
solution (plus/minus 34\%) from the distribution of 10000 simulations has a
slope and intercept of $4.39\pm0.35$ and $8.26\pm0.06$ respectively, and is
therefore consistent with our two other symmetrical regression analyses for
this data.

\subsection{All galaxies}

Figures~\ref{Fig3}b and \ref{Fig3}c include both the inactive and active
galaxies. 
A few galaxies stand out and may be worthy of identification. For example, 
the barred galaxy in the lower-left of Figure~\ref{Fig3}c (with $M_{\rm
  bh}=4\times10^5 M_{\odot}$ and $\sigma = 48$ km s$^{-1}$) is galaxy number 62
(see Table~\ref{Tab1}), a face-on Sc galaxy that may be barred. 
The four most discrepant non-barred AGN, found to the right of the shaded
region and with $M_{\rm bh} \sim 2\pm1 \times 10^7 M_{\odot}$ are galaxies 6,
8, 9 and 57 from Table~\ref{Tab1}.  Galaxy number 6 is poorly resolved, making
its morphology difficult to dissect. Galaxy number 8 is an edge-on disk galaxy
making it difficult to detect a bar if one exists.  The 'inactive' galaxy with a SMBH
mass of $\sim10^9 M_{\odot}$, which appears ten times greater than the value
expected from the relation, is NGC~5252.

We caution that Figures~\ref{Fig3}b and \ref{Fig3}c are probably not as strong
evidence against a separation of barred and unbarred galaxies as a first
inspection would suggest.  This is because many of the galaxies with velocity
dispersions less than 80-90 km s$^{-1}$ are too faint and/or not well enough
resolved for us to identify if a bar is present.  However, most galaxies which
deviate from the ``barless'' $M_{\rm bh}$-$\sigma$ relation,
in the sense that they have overly-large velocity dispersions, do tend to be
barred galaxies.  If we have failed to identify bars in the sample with
$\sigma > 85$ km s$^{-1}$, then correcting this would only increase the
population of outlying galaxies that are barred.
Among the 68 AGN with $\sigma > 85$ km s$^{-1}$, 38 are barred or show some
indication of possibly being barred.  This equates to 56\% of a sample which
includes both elliptical and disk galaxies.
Only 4 of the 30 non-barred AGN galaxies with $\sigma > 85$ km s$^{-1}$ 
display a clear departure to the right of the barless $M_{\rm bh}$-$\sigma$
relation, while roughly half of the 38 barred AGN galaxies are deviant in this
manner.

It turns out that this behaviour could have been predicted prior to the
confirmation provided here.  This is because it was known that 
a) large scale stellar bars are much more common in 
narrow line Seyfert 1 galaxies than broad-line Seyfert 1 galaxies
(Crenshaw et al.\ 2003; Deo et al.\ 2006) and that b) narrow-line Seyfert 1
galaxies are observed to reside rightward of the $M_{\rm bh}$-$\sigma$
relation traced by broad-line Seyfert 1 galaxies 
(Mathur et al.\ 2001; Wandel 2002; Grupe \& Mathur 2004; Mathur \& Grupe 2005b). 
% but see Komossa \& Xu 2007). 

For the 82 non-barred galaxies --- taken from the inactive sample of 50
galaxies plus the sample of 88 active galaxies --- the (symmetrical) 
bisector BCES regression analysis gives the relation 
\begin{equation}
\log(M_{\rm bh}/M_{\odot}) =
(8.18\pm0.05) + (4.05\pm0.18)\log [\sigma/200\, {\rm km\, s}^{-1}], 
\end{equation}
with an r.m.s.\ scatter of 0.42 dex in the $\log M_{\rm bh}$ direction.
The slope of this relation is shallower than the (symmetrical)
$M_{\rm bh}$-$\sigma$ relations presented by Ferrarese \& Merritt (2000) and 
Ferrarese \& Ford (2005), the latter reporting a slope of $4.86\pm0.43$. 
Obviously from Figure~\ref{Fig3}, the exclusion of barred galaxies results in
the shallower slope reported here.  A slope of 4.24 is obtained when we 
include all the galaxies.

Using the (non-symmetrical) ordinary least squares regression within the BCES routine 
which minimises the scatter in the $\log M_{\rm bh}$ direction, 
and is therefore preferred for predicting SMBH masses, one obtains the
expression 
\begin{equation}
\log(M_{\rm bh}/M_{\odot}) =
(8.15\pm0.05) + (3.89\pm0.18)\log [\sigma/200\, {\rm km\, s}^{-1}]. 
\label{Eq_non}
\end{equation}
This regression does however yield a comparable r.m.s.\ scatter of 0.41 dex 
in the $\log M_{\rm bh}$ direction.  Using the two other routines from
Section~\ref{Inactive} produces a consistent result.  While this ``barless'' relation is 
also consistent with the previous (non-symmetrical) $M_{\rm bh}$-$\sigma$ relations
presented by Gebhardt et al.\ (2000) and Tremaine et al.\ (2002), the 
latter reporting a slope of $4.02\pm0.32$, 
equation~\ref{Eq_non} has been derived using 82 non-barred galaxies rather than 31
barred and barless galaxies and is not applicable to barred 
galaxies.  While inclusion of the barred galaxies results in the relation
$
(8.03\pm0.05) + (3.94\pm0.19)\log [\sigma/200\, {\rm km\, s}^{-1}]$, with an 
r.m.s.\ scatter of 0.47 dex, this overlooks the fact that the scatter is not
uniformly distributed about this relation. 
A more reliable estimate of the SMBH masses in barred galaxies would come from 
a subtraction of 0.3 dex from the barless $M_{\rm bh}$-$\sigma$ relation's
prediction, i.e. dividing by two, and assigning an unceratinty of $\sim$0.4 dex.

\subsection{Third parameters}

The presence of a bar can, but does not necessarily, result in a galaxy
residing significantly off the barless $M_{\rm bh}$-$\sigma$ relation.  Given
that, at a fixed velocity dispersion, the range in SMBH mass spans a factor of
$\sim$50 when considering both non-barred and barred galaxies, it is obviously
neither desirable nor optimal to use a single $M_{\rm bh}$-$\sigma$ relation
for all galaxy types.  The predictive power of such a relation would be too
weak for practical purposes.

While the ``barless $M_{\rm bh}$-$\sigma$ relation'' is useful for non-barred
galaxies, in order to accommodate the barred galaxies it would be advantageous
if one could identify a third parameter which could account for their offset
nature.  While barred galaxies do have smaller bulges than large elliptical
galaxies, the use of effetcive half-light bulge radii as a third parameter may
only serve to complicate matters.  This is because non-barred galaxies with
equally small bulge sizes do not deviate from the locus of points defining the
barless $M_{\rm bh}$-$\sigma$ relation and thus bulge size is not the reason why the
barred galaxies appear offset.  Obviously though, plotting the $M_{\rm
  bh}$-$\sigma$ residuals against bulge size will yield a trend, and this has
led some to (perhaps prematurely) 
conclude that a SMBH ``fundamental plane'' involving $M_{\rm bh},
\sigma$ and $R_{\rm e}$ exists (see Feoli \& Mele 2007;
Aller \& Richstone 2007; Hopkins et al.\ 2007).  Nonetheless, if
the virial expression $M_{\rm bulge} \propto \sigma^2 R_{\rm e}$ roughly
produces bulge masses, and $M_{\rm bh}$ is related to $M_{\rm bulge}$, then
one should expect the addition of $R_{\rm e}$, as initially done by Marconi \&
Hunt (2003), to make sense.  However for pure elliptical galaxies the addition
of $R_{\rm e}$ appears to offer no benefit (Graham 2008a) and thus further
undermines the value of $R_{\rm e}$ as a potential third parameter.  We feel
that more galaxy structural data may be needed to resolve this issue.

Graham (2008a) speculated that the offset nature of the barred galaxies 
may be related to an increased 
exposure to the radial orbits of stars in bars.  If this is the case, then one
would expect that the bars in more inclined (i.e., edge-on) galaxies will, in
general, have increased velocity dispersions if the bars are orientated toward
us. Of course an edge-on galaxy may still have a bar which is aligned along
the plane of the sky, yielding no increase to the observed velocity
dispersion.  Nonetheless, to probe this idea we can use the inclination of the
disks, traced by the ellipticity of their outer isophotes.  The inclinations,
$i$, of the disk galaxies are of course related to their observed
ellipticities via $\cos(i) = (1 - \epsilon)$, which were themselves obtained
from the IRAF routine ELLIPSE and are listed in Table~1.  The data in
Figure~\ref{Fig4}, which shows the vertical residuals about the ``barless
$M_{\rm bh}$-$\sigma$'' for the barred galaxies, plotted against the apparent
disk ellipticity ($\epsilon$), are not particularly supportive of this
scenario as they do not reveal a distribution of residuals which broadens at
larger elliptcities.

The near uniform distribution of points in Figure~\ref{Fig4} suggests that
something other than disk inclination may be at play.  As noted by Gadotti \&
Kauffmann (2009), it could be the age of the bar such that older bars have
had time to increase the central velocity dispersion and thereby drive
galaxies off the barless $M_{\rm bh}$-$\sigma$ relation.  Indeed, Perez et
al.\ (2009) have reported that older bars have greater velocity dispersions
than younger bars.  They also note that older bars have positive metallicity
gradients.  Unfortunately, such diagnosis requires spatially resolved spectral
information which is not trivial to acquire, and therefore difficult to
implement as a practical third parameter to reduce the scatter in the $M_{\rm
  bh}$-$\sigma$ diagram.

\section{Summary and Commentary}

\begin{figure}
\includegraphics[angle=270,scale=0.685]{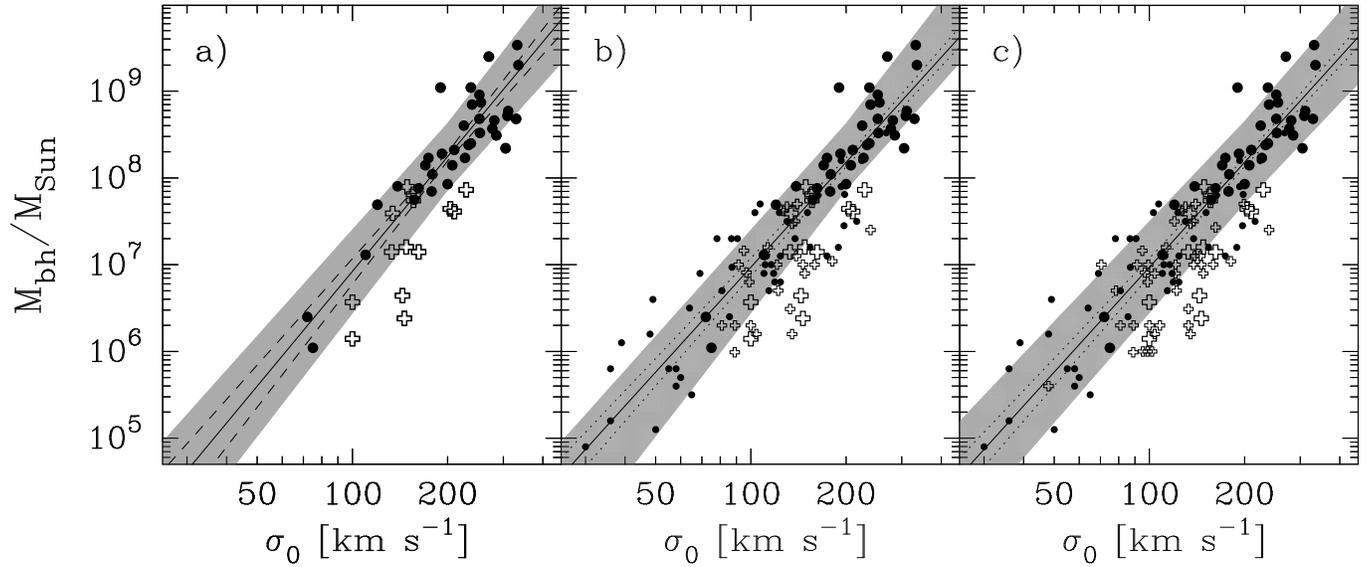}
\caption{
$M_{\rm bh}$-$\sigma$ diagram. Crosses denote galaxies with (suspected) bars.
Panel a) shows 50 galaxies with direct SMBH mass measurements, as tabulated by
Graham (2008b, his Table~1). 
The solid line shows the (symmetrical) linear regression to the 36 non-barred 
galaxies, while the dashed lines delineate the 1$\sigma$ uncertainty for this relation.
The shaded area extends this boundary by 0.33 dex in the $\log M_{\rm bh}$
direction.  The outlier at the high-mass end with $M_{\rm bh} = 10^9
M_{\odot}$ is NGC~5252,
Panel b) includes the AGN from Table~\ref{Tab1} which are designated as either
having a bar or not having a bar.  While the shaded region has been kept the
same as in panel a) for comparison purposes, the solid line shows the (symmetrical) linear
regression to the 82 non-barred galaxies, while the dotted lines delineate the
1$\sigma$ uncertainty for this relation.
Panel c) additionally includes those AGN for which there may be a bar; that
is, systems designated as ``maybe'' are treated as ``yes'' in this panel (see
Table~\ref{Tab1}).  The shaded region extends the boundary around the dotted
lines by 0.42 dex --- which is the r.m.s.\ scatter about the barless $M_{\rm
 bh}$-$\sigma$ relation for the 82 non-barred (AGN and inactive) galaxies.
}
\label{Fig3}
\end{figure}

\begin{figure}
\includegraphics[angle=270,scale=0.45]{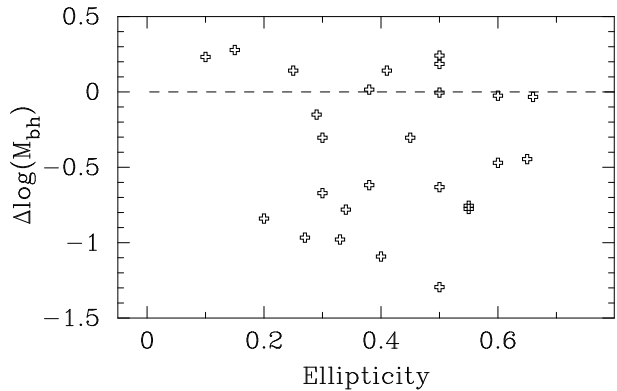}
\caption{
Vertical residuals of the barred galaxies 
about the barless $M_{\rm bh}$-${\sigma}$ relation 
shown in Figure~\ref{Fig3}b plotted against the ellipticity
of each galaxy's outer isophotes --- a measure of their disk's inclination. 
}
\label{Fig4}
\end{figure}

A number of immediate observations can be made from the $M_{\rm bh}$-$\sigma$
diagram shown in Figure~\ref{Fig3}.
(1) While AGN have a reputation for displaying a greater level of scatter in
the $M_{\rm bh}$-$\sigma$ diagram than inactive galaxies, this is primarily 
because the AGN sample contains many barred galaxies.  That is, the greater 
uncertainty from (reverberation mapping and emission line)-derived black hole 
masses, compared to direct measurements, is not the main cause of the
increased scatter in the $M_{\rm bh}$-$\sigma$ diagram. 
(2) The barred galaxies, in the current sample of 138 galaxies, 
tend not to have SMBH masses greater than $10^8 M_{\odot}$. 
(This is in part a reflection on the maximum masses of bulges in disk galaxies.)
(3) To the left of the shaded region in Figure~\ref{Fig3}b, galaxies 
tend not to be barred, while to the right of the shaded region most of the galaxies are barred.
%
% (4) The barred galaxies also appear to have velocity dispersions greater than
% 80-90 km s$^{-1}$, making them stand out among the galaxies with $10^6 <
% M_{\rm bh}/M_{\odot} < 10^7$.  However, this may be a reflection of the
% difficulty in identifying bars in smaller galaxies.

This structure within the $M_{\rm bh}$-$\sigma$ plane inhibits the accuracy to
which black hole masses can be predicted in other (barred) galaxies for which only the
velocity dispersion is known.  While for elliptical and barless disk galaxies
one can still obtain an accurate estimate using the ``barless $M_{\rm
  bh}$-$\sigma$ relation'' (equation~2), the situation is less favourable for
the barred galaxies.  It may be that 
dynamically old bars have increased the central velocity dispersion in these
galaxies. 

The obvious consequence of this work is that the galaxy velocity dispersion is
not the sole driving force which determines the masses of black holes at the
centres of galaxies.
% The velocity dispersion of a galaxy, or its spheroidal component, {\it
% prior} to the formation of a bar might, however, still be the driving parameter.  
% although why this parameter should then cease to drive SMBH growth when
% a bar is present suggests that it may have never been the driving force.
It is also apparent that dividing one's sample into elliptical and disk galaxies
is not an appropriate approach to take as this would come at the expense of 
accurate SMBH masses for non-barred disc galaxies. 

Barred galaxies do not appear as outliers in current relations
involving black hole mass and spheroid luminosity (Graham
2008a, his figure~3; Bentz et al.\ 2009). 
Increasing the sample sizes used in these diagrams, which has currently been limited
to only 20 
to 30 objects, is recommended.  Acquiring deeper, higher-resolution, near-infrared images
than 2MASS would be a valuable step forward toward better understanding the
driving forces that dictate the masses of supermassive black holes at the
centres of galaxies. 

It should also be insightful to determine, in subsequent work, the scaling 
factor (e.g., Onken et al.\ 2004) which brings reverberation-based SMBH masses 
for nonbarred AGN into better agreement with the
barless $M_{\rm bh}$-$\sigma$ relation for quiescent galaxies.
The prevalence of barred AGN in such past analyses may have skewed these 
measurements and hindered constraints for the different broad-line region
models.
Ongoing and future observing reverberation mapping campaigns 
% e.g. There is a Lick AGN monitoring project for 13 AGN (Bentz et al.\ 2008,
% ApJ, 689, L21).
may therefore like to consider focussing on non-barred AGN for which the 
velocity dispersion is, or can be readily, acquired.

% \acknowledgements

\end{document}